\documentclass[twoside,twocolumn,9pt]{article}
\usepackage{extsizes}
\usepackage[super,sort&compress,comma]{natbib} 
\usepackage[version=3]{mhchem}
\usepackage[left=1.5cm, right=1.5cm, top=1.785cm, bottom=2.0cm]{geometry}
\usepackage{balance}
\usepackage{mathptmx}
\usepackage{sectsty}
\usepackage{graphicx} 
\usepackage{lastpage}
\usepackage[format=plain,justification=justified,singlelinecheck=false,font={stretch=1.125,small,sf},labelfont=bf,labelsep=space]{caption}
\usepackage{float}
\usepackage{fancyhdr}
\usepackage{fnpos}
\usepackage[english]{babel}
\addto{\captionsenglish}{%
  
}
\usepackage{array}
\usepackage{droidsans}
\usepackage{charter}
\usepackage[T1]{fontenc}
\usepackage[usenames,dvipsnames]{xcolor}
\usepackage{setspace}
\usepackage[compact]{titlesec}
\usepackage{hyperref}

\usepackage{epstopdf}

\definecolor{cream}{RGB}{222,217,201}

\begin{document}

\pagestyle{fancy}
\thispagestyle{plain}
\fancypagestyle{plain}{
\renewcommand{\headrulewidth}{0pt}
}

\makeFNbottom
\makeatletter
\renewcommand\LARGE{\@setfontsize\LARGE{15pt}{17}}
\renewcommand\Large{\@setfontsize\Large{12pt}{14}}
\renewcommand\large{\@setfontsize\large{10pt}{12}}
\renewcommand\footnotesize{\@setfontsize\footnotesize{7pt}{10}}
\makeatother

\renewcommand{\thefootnote}{\fnsymbol{footnote}}
\renewcommand\footnoterule{\vspace*{1pt}%
\color{cream}\hrule width 3.5in height 0.4pt \color{black}\vspace*{5pt}} 
\setcounter{secnumdepth}{5}

\makeatletter 
\renewcommand\@biblabel[1]{#1}            
\renewcommand\@makefntext[1]%
{\noindent\makebox[0pt][r]{\@thefnmark\,}#1}
\makeatother 
\renewcommand{\figurename}{\small{Fig.}~}
\sectionfont{\sffamily\Large}
\subsectionfont{\normalsize}
\subsubsectionfont{\bf}
\setstretch{1.125} 
\setlength{\skip\footins}{0.8cm}
\setlength{\footnotesep}{0.25cm}
\setlength{\jot}{10pt}
\titlespacing*{\section}{0pt}{4pt}{4pt}
\titlespacing*{\subsection}{0pt}{15pt}{1pt}

\fancyfoot{}
\fancyfoot[LO,RE]{\vspace{-7.1pt}\includegraphics[height=9pt]{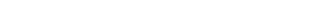}}
\fancyfoot[CO]{\vspace{-7.1pt}\hspace{11.9cm}\includegraphics{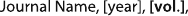}}
\fancyfoot[CE]{\vspace{-7.2pt}\hspace{-13.2cm}\includegraphics{head_foot/RF}}
\fancyfoot[RO]{\footnotesize{\sffamily{1--\pageref{LastPage} ~\textbar  \hspace{2pt}\thepage}}}
\fancyfoot[LE]{\footnotesize{\sffamily{\thepage~\textbar\hspace{4.65cm} 1--\pageref{LastPage}}}}
\fancyhead{}
\renewcommand{\headrulewidth}{0pt} 
\renewcommand{\footrulewidth}{0pt}
\setlength{\arrayrulewidth}{1pt}
\setlength{\columnsep}{6.5mm}
\setlength\bibsep{1pt}

\makeatletter 
\newlength{\figrulesep} 
\setlength{\figrulesep}{0.5\textfloatsep} 

\newcommand{\topfigrule}{\vspace*{-1pt}%
\noindent{\color{cream}\rule[-\figrulesep]{\columnwidth}{1.5pt}} }

\newcommand{\botfigrule}{\vspace*{-2pt}%
\noindent{\color{cream}\rule[\figrulesep]{\columnwidth}{1.5pt}} }

\newcommand{\dblfigrule}{\vspace*{-1pt}%
\noindent{\color{cream}\rule[-\figrulesep]{\textwidth}{1.5pt}} }

\makeatother

\twocolumn[
  \begin{@twocolumnfalse}
{\includegraphics[height=30pt]{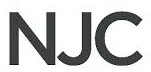}\hfill\raisebox{0pt}[0pt][0pt]{\includegraphics[height=55pt]{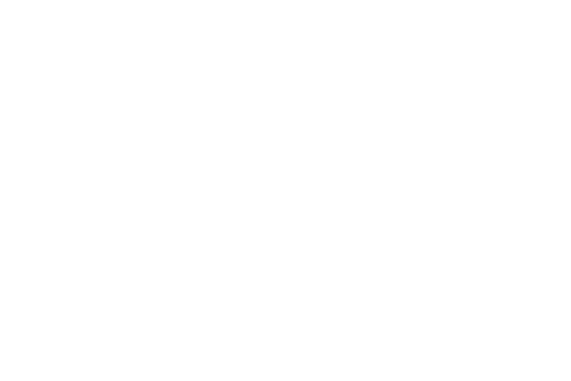}}\\[1ex]
\includegraphics[width=18.5cm]{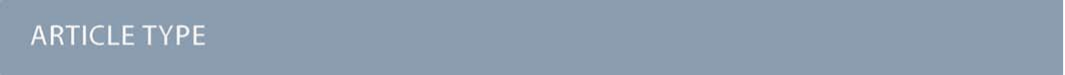}}\par
\vspace{1em}
\sffamily
\begin{tabular}{m{4.5cm} p{13.5cm} }

\includegraphics{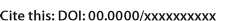} & \noindent\LARGE{\textbf{A comprehensive study of electronic and piezoelectric properties of Li-based Tin-halide perovskites from GGA and Meta-GGA}} \\
\vspace{0.3cm} & \vspace{0.3cm} \\

 & \noindent\large{Celestine Lalengmawia,$^{a}$\textit{$^{b}$} Zosiamliana Renthlei,$^{a}$\textit{$^{b}$} Shivraj Gurung,$^{b}$ Lalhriat Zuala,$^{b}$ Lalrinthara Pachuau,$^{b}$ Ningthoujam Surajkumar Singh,$^{b}$ Lalmuanpuia Vanchhawng,$^{b}$ Karthik Gopi,\textit{$^{c}$} A. Yvaz,\textit{$^{d}$} and D. P. Rai\textit{$^{a}$}{$^\ast$}} \\

\includegraphics{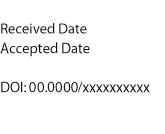} & \noindent\normalsize{	Wide bandgap semiconductors (WBGs) are predicted to be the potential materials for energy generation and storing. In this work, we used density functional theory (DFT) that incorporates generalized gradient approximation (GGA) and meta-generalized gradient approximation (mGGA) methods to explore the various properties of the LiSnCl$_3$ and LiSnBr$_3$ perovskites. The structural stabilities, charge transfer, electronic, optical, mechanical, and piezoelectric properties are studied. Herein, we report that these rarely studied materials are eco-friendly and look promising for optoelectronics and piezoelectric applications.} \
\end{tabular}

 \end{@twocolumnfalse} \vspace{0.6cm}

  ]

\renewcommand*\rmdefault{bch}\normalfont\upshape
\rmfamily
\section*{}
\vspace{-1cm}


\footnotetext{\textit{$^{a}$Advanced Computation of Functional Materials Research Lab (ACFMRL) Department of Physics, Mizoram University, Aizawl-796004, India}}
\footnotetext{\textit{$^{b}$Physical Sciences Research Center (PSRC), Department of Physics, Pachhunga University College,  Aizawl-796001, India}}

\footnotetext{\textit{$^{c}$Department of Nuclear Physics, University of Madras, Guindy Campus, chennai--600025, India}}
\footnotetext{\textit{$^{d}$World-class research center "Advanced Digital Technologies", State Marine Technical University, Saint Petersburg, 190121 Russia}}
\footnotetext{{$^\ast$}Email:\textcolor{blue}{dibyaprakashrai@gmail.com}}
\footnotetext{orcid:https://orcid.org/0000-0002-3803-8923}

\footnotetext{\dag~Supplementary Information available: [details of any supplementary information available should be included here]. See DOI: 10.1039/cXCP00000x/}

\footnotetext{\ddag~Additional footnotes to the title and authors can be included \textit{e.g.}\ `Present address:' or `These authors contributed equally to this work' as above using the symbols: \ddag, \textsection, and \P. Please place the appropriate symbol next to the author's name and include a \texttt{\textbackslash footnotetext} entry in the the correct place in the list.}


\section{Introduction}
To meet the current energy crisis, harvesting clean energy has been a primary objective in the contemporary world. \cite{Yu2019b,Yang2020c} Researchers, from past to present, have work on materials which are efficient for energy-harvesting. The halide-based perovskites are known for energy-efficient materials due to their ability to store energy and high conversion rate.\cite{Kim2017a} These materials are extensively studied owing to their distinctive physical characteristics and  economical production. The other merit includes their ability to tune the electronic band gap, defect tolerance, long-range carrier diffusion lengths, and high piezoelectric coefficients, which makes them promising for various technological applications. \cite{Bowen2014a}\cite{Park2016a}\cite{Das2022b} Several techniques have been implemented to extract and store energy in which solar energy is the most common and regarded as the most promising due to its inexhaustible source. \cite{Sutherland2016}\cite{Chen2015}\cite{Jena2019a} The development of highly efficient solar-cell devices by utilizing the halide perovskites in the solar panels for absorbing large range of solar-spectra have gathered further research interest and regarded as the clean source of renewable energy. \cite{Liu2022a, ThiHan2022a, Hasan2022a, Rahman2022a} Despite, being the ideal and clean method of producing energy, solar panels also have some limitations.\cite{Rajagopal2017} Such as the low efficiency for mass utilization, occupies large space, unable to store energy, the electrode contaminated due to the exposure to environment, disposal of solar panel is still a big issue.\cite{Rahaman2018a} \cite{Rahaman2018a, Schwarz1996c, Mustafa2023a} 

\par Apart from solar panels, there are numerous ways of generating energy. Some of them are through- Thermal energy generation\cite{LUO2015511} which requires burning of biogas or natural gas, where the burning heat can generate electricity. But this type of power generation exploits environment and causes air pollution. Secondly, Coal energy generation,\cite{GOMEZCALVET2021233} which needs a complex method of mining and processing techniques. Furthermore, a nuclear power generation,\cite{NAKAGAWA2022132530} which uses the spiltting of Uranium-atom to release and produce clean energy. But except for its reactor, which works in the same way with Thermal energy generator and could be hazardous to human and environment. In addition, we have also garnered power through the wind\cite{IBRAHIM2011815} and hydropower\cite{AZAD20202230} generations, these generators depend on the external pressures of wind and water to generate energy. Among these various techniques of energy generation, piezoelectricity is one of them, in which the materials converts mechanical pressure into electrical energy, which seems to be a greener approach.\cite{Hao2019a, Park2014a, Bowen2014b, Park2016b} The energy harvesting technique from piezoelectric material looks more safer as they are independent of the external factors like change in weather condition or air pollution etc. \cite{Celestine2024b, Jeong2017a} Other than energy production and storage, the piezoelectricity have rigorous applications in sensor devices like biomedical and nanorobot.\cite{Xue2021a, Bouhmaidi2022a}

\par Majority of non-centrosymmetric wide band gap materials are more prone to piezoelectric phenomena.\cite{ParrydeepKaurSachdeva2015} So far, several piezoelectric materials have been discovered and most common ones are zinc sulfide (ZnS)\cite{Saksena1951} which yields the piezoelectric coefficient of 0.161 Cm$^{-2}$,\cite{FERAHTIA201488}silicon carbide (SiC)\cite{Kudimi2012} having reported its piezo-cofficient of 0.323 Cm$^{-2}$\cite{Cimalla_2007} and lead zirconate titanate (PZT)\cite{Zhang2017} with -4.7 Cm$^{-2}$ piezoelectric response.\cite{CATTAN199960} Among all the studied piezoelectric materials PZT exhibit high piezoelectric response but its utilization is limited due to the presence of toxic Pb element. To overcome this issue, there has been an enormous effort in replacing the Pb element without compromising the efficiency.  \cite{Park2020b} Inorganic metal halide perovskites have shown lots of potential owing to their non-centrosymmetric crystal phase, high dielectric polarization, long-range carrier diffusion, anisotropic absorption coefficients, defect tolerance, tunable bandgap, high piezoelectric coefficient, large exciton binding energy, and high stability.\cite{Celestine2024c} Inorganic metal halide perovskites have also shown promising characteristics for their direct applications in optoelectronics to photocatalysis. In recent years, a large number of flexible lead-free piezoelectric materials have been synthesised and investigated in the lab by utilizing the solid-state process and time-dependent study to obtained the materials and measured the piezoelectric properties.\cite{MADHUBABU2021138560} By electrospinning talc/polyvinylidene fluoride (PVDF) nanocomposites, Anandhan et al.\cite{D0SM00341G} synthesized a polymer  for flexible piezoelectric nanogenerator. Piezoelectric sensors have garnered a lot of interest for their possible applications in human physiological signal monitoring and wearable energy harvesting.\cite{ElKacimi2018} \cite{Luo2020}\cite{Toyabur2017}\cite{Min2023}. 

\par For our investigation we have  selected metal-halide perovskites having a chemical formula ABX$_3$, where the monovalent halogen atoms are the X-site anions, the divalent atoms are the B-site cations, and the monovalent alkali metals make up the A-site cations.\cite{Rao2003}\cite{Sahoo2018} Here, we used Lithium (Li$^-$) as the A-site atoms, with Tin (Sn$^{2+}$) at B-site and the X-site is occupied by the halogens (Cl$^-$, and Br$^-$).
Ghani et al.\cite{Ghani2024} have reported the structural stability of LiSnX$_3$ from the first-principles calculation. The proposed compounds being noncentrosymmetrically stable  and non-toxic (Pb-free) material, further research have also been performed for their potential applications in the energy storage system (battery). \cite{Park2010}\cite{Loftager2016}\cite{Nzereogu2022} Herein, we have calculate and report the structural stabilities, electronic profiles, optical and piezoelectric properties of the metal halide-based perovskites LiSnX$_3$.

\section{Computational Details}
All the computations in this work were done using QuantumATK Q-2019.12's implementation of density functional theory (DFT)\cite{Nityananda2017}, which relies on the linear combination of the atomic orbital (LCAO) approach.\cite{Smidstrup2019a} For electron-ion interactions, we have employed two types of exchange-correlations namely: generalized gradient approximation (GGA)\cite{Carmona-Espindola2015}\cite{Perdew1996c} within the course of Perdew-Burke-Ernzerhof (PBE) functional\cite{Perdew1997} and meta-generalized gradient approximation (mGGA)\cite{Sun2011}\cite{Karasiev2022} which uses SCAN exchange-correlation functional.\cite{Yao2017} And to fully achieve the cell optimization, we adopted the Limited-memory Broyden-Fletcher-Goldfarb-Shanno (LBFGS) optimization algorithm rooted from the  quasi-Newton methods,\cite{Zhao2021} which computes an inverse hessian matrix whose system suits well for optimizing problems with finite variables. We have set the parameters of the unit cell volumes, space groups, and atomic positions to be relaxed throughout the geometry optimization. In order to ascertain the structural and cell energy convergence, the criterion for Hellmann-Feynmann force and stress tolerance were set to 0.01 eV \AA$^{-1}$ and 0.00001 eV \AA$^{-3}$, respectively, with a maximum step size of 0.2 \AA and a maximum number of 200 convergence steps.

\par An occupation method of Fermi-Dirac has been chosen for calculating the numerical accuracy with the broadening energy and density mesh cut-off at 25 meV and 85 Hatree, respectively. For each atom- Li, Sn, Cl and Br, a medium basis set of PseudoDojo potential, which is equivalent to the double zeta polarized (DZP), have been adopted.\cite{VanSetten2018a} A Hamiltonian mixing variable with a tolerance factor of 0.0001 at 0.1 damping factor under the Pulaymixer algorithm have been set up to control the iteration. Utilizing the Monkhorst-Pack scheme,\cite{Monkhorst1976a} we have sampled a 10$\times$10$\times$10 k-mesh points for geometry optimization while for property calculations a denser sampling of 12$\times$12$\times$12 with the same convergence basis have been adopted.

\section{Results and Discussion}
\subsection{Structural Properties}
\begin{figure}[hbt!]
	\centering
	\includegraphics[height=5.5cm]{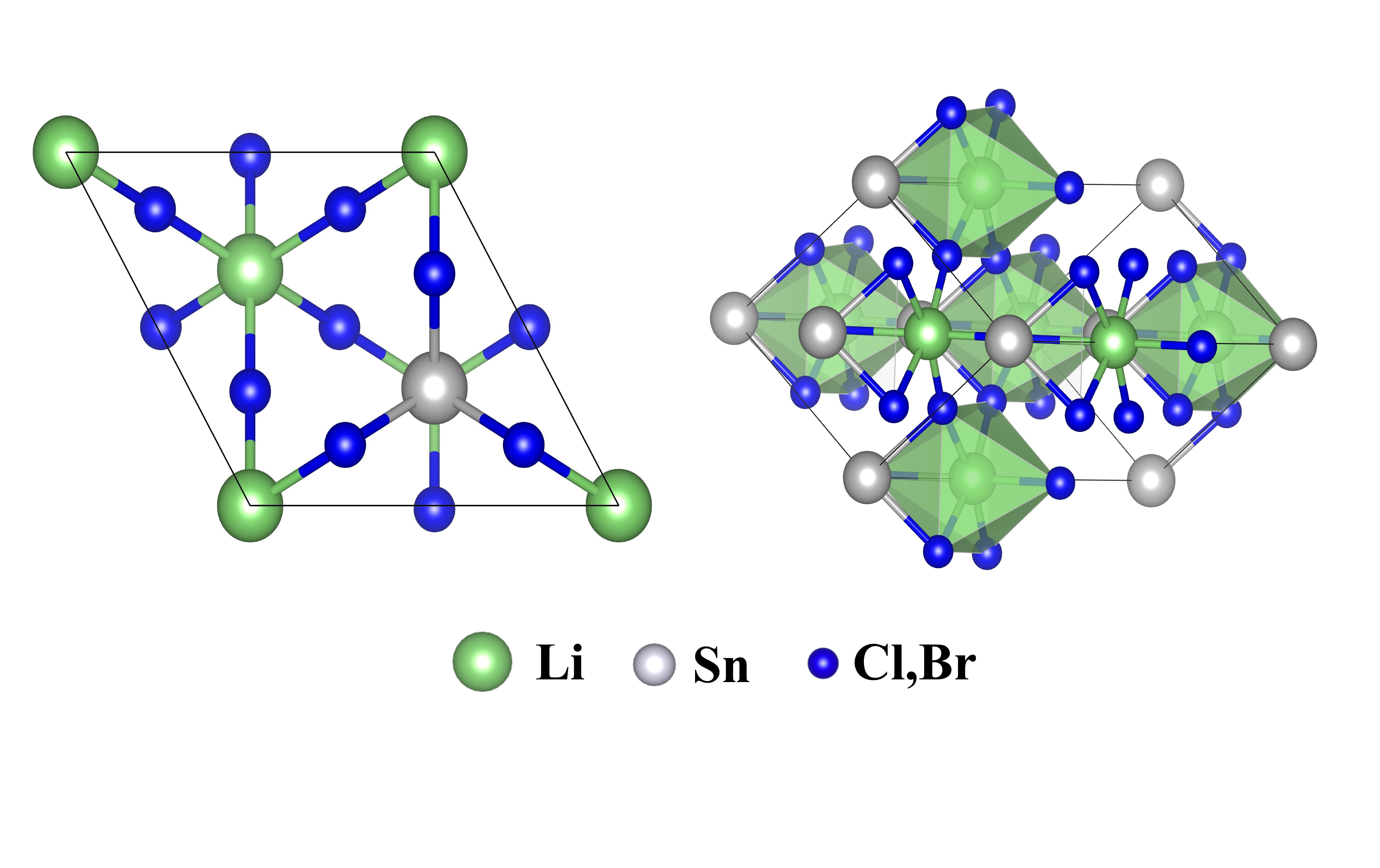}
	\caption{Schematic presentation of 2D and 3D crystal structure of LiSnX$_3$(X=Cl and Br) using VESTA program \cite{Momma2008a} }
	\label{Fig.1 structural}
\end{figure}

The under studied halide perovskites LiSnX$_3$ (X = Cl, and Br) have been observed to exhibit a trigonal symmetry of R3m (160) space group at their pristine states. (See Fig.\ref{Fig.1 structural}) The optimized parameters of LiSnCl$_3$ and LiSnBr$_3$ employing GGA and mGGA functionals are provided in Table \ref{Table 2}. The role played by the employed functionals have compared with some of the existing theoretical datas of LiSnCl$_3$. In their primitive cells, there are numbers of 15 atoms consisting of three Li-atoms, three Sn-atoms, and nine Cl/Br-atoms, respectively. The atomic positions of A-site (Li-atom), B-site (Sn-atom) and halogens (Cl/Br-atoms) are taken by using the Wyckoff coordinates.\cite{Hasan2022c}

\begin{table}[hbt!]
	\small
	\caption{\ Calculated ionic radii (\textit{r}) in \AA and octahedral factor (\textit{$\mu$}) of LiSnX$_3$ (X = Cl, and Br)} 
	\begin{tabular*}{0.48\textwidth}{@{\extracolsep{\fill}}l|l|l|l}
		\hline
		Atom(s) &  Compound(s) & \textit{$\mu$}& Other's work\\
		\hline
		Li$_{(A)}$       & LiSnCl$_3$ & 0.73&0.65\cite{Tanaka2018} \\
		Sn$_{(B)}$       & & &\\
		Cl$_{(X_{1})}$    & LiSnBr$_3$ & 0.67& 0.60\cite{Tanaka2018}\\
		Br$_{(X_{2})}$   & & &\\
		\hline
	\end{tabular*}
	\label{Table 1}
\end{table} 

\begin{table*}[hbt!]
	\caption{ Calculated lattice constants (a=b,c), volumes (\AA), and bandgaps (\textit{E$_g$})  of LiSnX$_3$ (X = Cl, and Br)} 
	\begin{tabular*}{\textwidth}{@{\extracolsep{\fill}}l|l|l|l|l|l}
		\hline
		Compound(s) &Functionals& a=b,c &V (\AA$^3$) & \textit{E$_g$} (eV)&  Other's work (\textit{E$_g$}) \\
		\hline			
		LiSnCl${_3}$     & GGA  & 7.11, 9.34 & 409.5 & 3.81&1.35, 3.38, 4.94\cite{Ghani2024,Park2019,Lamichhane}\\
		LiSnBr${_3}$    &  GGA  & 7.58, 10.12& 504.9&3.18& ---\\
		LiSnCl${_3}$   & mGGA  & 6.98, 9.18&388.1& 4.17&---\\ 
		LiSnBr${_3}$  & mGGA& 7.58, 10.12&505& 3.67& ---\\
		\hline
	\end{tabular*}
	\label{Table 2}
\end{table*} 

\par Since there have been rarely reported data for these systems. For calculating the structural stabilities, we have the ionic radii of each atoms as presented in Table \ref{Table 1}. Using the values of the ionic radii, the octahedral or $\mu$ factor can be calculated. This can be expressed as\cite{Ji2019}
\begin{equation}
\mu=\dfrac{R_{B}}{R_{X}}
\end{equation}
\par The calculated octahedral or $\mu$ factors are listed in Table \ref{Table 1}. The employed empirical scheme- octahedral factors do not considered the chemical interplay between the integral parts of the elements and ions. For that reason, the role played by each individual atoms solely depends on the stability of the perovskite systems. So, to form a halide perovskite crystal empirically, it would have the octahedral factor, $\mu$ from 0.442 to 0.895, respectively. From Table \ref{Table 1}, we have found that the investigated systems are stable as the calculated $\mu$ factor fulfill the octahedral stability criteria.

\par Utilizing GGA and mGGA, it can be seen from Table \ref{Table 2} the lattice constants increases from Cl$\rightarrow$ Br attributed to the larger atomic orbitals, eventually reduce the inter-atomic distances between them. However, for LiSnBr$_3$, that the lattice constants are found to be the same within both the functionals. The \textit{E$_g$} of ABX$_3$ (X=Cl) compounds shows wider bandgaps than that of compounds with the Br-atoms owing to the higher electronegativity of Cl-atom, facilitate stronger orbital-hybridization. Our results of wide bandgap (WBG) are compared with some of the existing data (comparative results are tabulated in in Table \ref{Table 2}).  From our literature survey, we have found that the wide bandgaps (WBGs) Li-based semiconductors  are promising materials for battery or power capacitors crucial for storing energy for further used.\cite{Wellmann2021}\cite{Kumar2022} The use of WBGs halide perovskites make devices lighter in weight, flexible, cheaper in fabrication. Other than that halide perovskites are in high demand due to the faster electrons transfer plausible for electronic device integration.\cite{Bhattacharya2018}

\subsection{Mulliken Population Analysis}
To understand the inter-atomic bonding within the compounds, we have utilized the mulliken population analysis(MPA)\cite{Mulliken1955} which provides an insight of bonding between atoms and charge distributions among the atoms within the materials. Herein, Table \ref{Table 3} we report the analyzed populations, charges and bond lengths of the studied halide perovskites-LiSnCl$_3$ and LiSnBr$_3$ using GGA and mGGA. The contribution of all atoms are given in terms of their charges of s- and p-orbitals. The presence of negative charge values in Cl/Br atom is a result of charge accumulation which have a direct indication of charge transfer from Sn/Li$\rightarrow$Br/Cl, creating charge depletion around Sn/Li atoms.  The results of charge depletion (+) and charge accumulation (-) by using GGA and mGGA are given in Table \ref{Table 3} .
\begin{table*}[hbt!]
	\caption{Calculated the atomic populations, mulliken charge and bond lengths of LiSnX$_3$ (X = Cl, and Br) using GGA and mGGA} 
	\begin{tabular*}{\textwidth}{@{\extracolsep{\fill}}l|l|l|l|l|l|l}
		\hline
		Compound(s)& Functionals&Species& Atomic population &Mulliken charge \textit{(e)} & Bond&  Bond length (\AA)\\
		&& &Total \ \ \ \ \ s \ \ \ \ \ \ p&&&\\
		\hline			
		LiSnCl${_3}$    &GGA & Li  & \ 2.39, 2.06, \ \ 0.32 &  (+)0.61 & Li-Sn&3.41\\
		&&&&&Li-Cl&2.67\\
		&&  Sn  & 13.35, 0.15, 11.41&  (+)0.65 &Sn-Cl&2.63\\
		&& Cl  & \ 7.41, \ 1.93, \ 5.48&(-)0.41&Cl-Cl &3.46\\ 
		&&&&&&\\
		LiSnBr${_3}$ & GGA &Li& \ 2.51, 2.15, \ 0.37&\ (+)0.49&Li-Sn& 3.62\\
		&&&&&Li-Br&2.85\\
		&& Sn& 13.49, 1.95, 11.54&  (+)0.51&Sn-Br & 2.77\\
		&& Br& \ \ 7.33, 1.94, \ 5.39&(-)0.33& Br-Br& 3.72\\
		\hline
		LiSnCl${_3}$  &mGGA& Li&\ \  2.53, 2.15, 0.38 & \ (+)0.47& Li-Sn& 3.98\\
		&&&&&Li-Cl&2.44\\
		&& Sn& \ 13.38, 1.90, 11.48&  (+)0.62& Sn-Cl& 2.58\\
		&& Cl& \ \ 7.37, 1.90, \ \ 5.49&(-)0.37&Cl-Cl & 3.39\\
		&&&&&&\\ 
		LiSnBr${_3}$  &mGGA& Li& \ \ 2.56, 2.21, \ \ 0.35& \ (+)0.44&Li-Sn & 3.59\\
		&&&&&Li-Br&2.69\\
		&& Sn& \ 13.46, 1.92, 11.55&  (+)0.54& Sn-Br& 2.73\\
		&& Br& \ \ \ 7.32, 1.93, \ 5.39&(-)0.32&Br-Br & 3.70\\
		\hline
	\end{tabular*}
	\label{Table 3}
\end{table*}
\subsection{Electron Localization Function}
\begin{figure*}[hbt!]
	\centering
	\includegraphics[height=7cm, width=8cm]{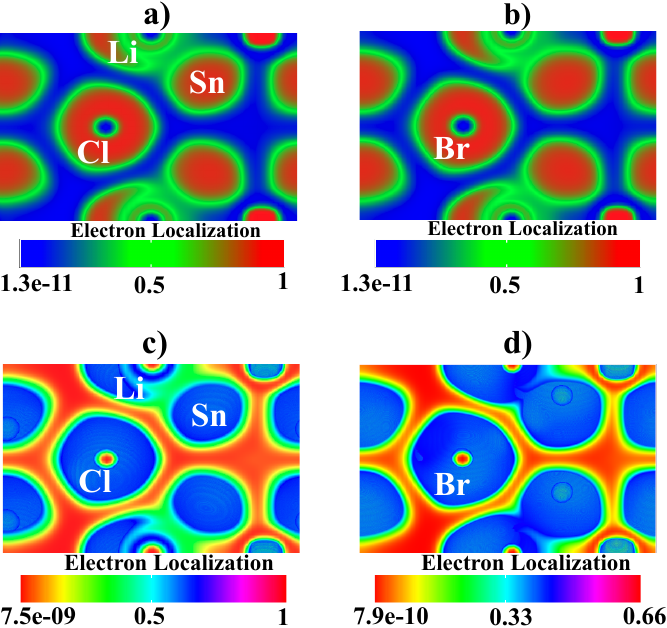}
	\includegraphics[height=7cm, width=8cm]{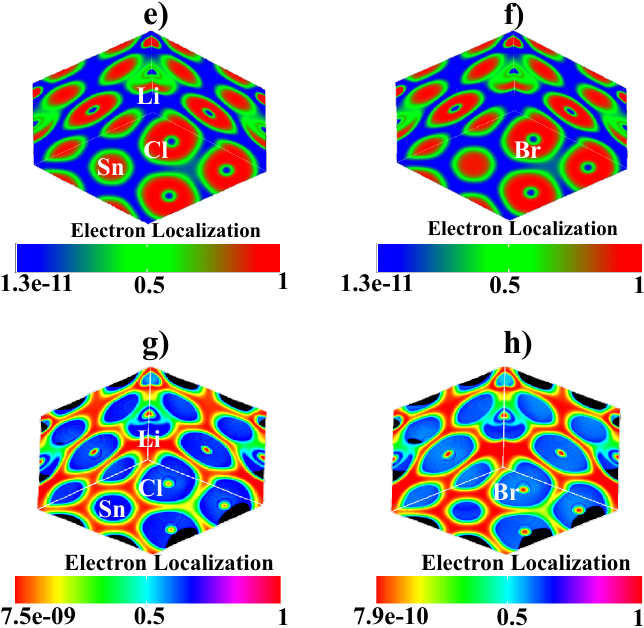}
	\caption{Schematic presentation of 2D and 3D electron localization function (ELF) of LiSnX$_3$(X=Cl and Br). Fig.\ref{Fig.2 ELF} (a-b, e-f) are obtained using GGA and Fig.\ref{Fig.2 ELF} (c-d, g-h) are obtained using mGGA}
	\label{Fig.2 ELF}
\end{figure*}

The electron localization function (ELF)\cite{Savin1997} has become an effective tool for qualitatively comprehending the behavior of electrons within the atomic system during the course of crystalline formation. A wide range of bonding scenarios can be explained, from the most common covalent to the metallic bond. The ELF does not solely depend on the employed basis sets and computational techniques. ELF plays a crucial role in understanding the chemical bond formation providing vital information about the electrons localization. We have displayed the 2D and 3D ELF maps of LiSnCl$_3$ and LiSnBr$_3$ employing GGA and mGGA in Fig.(\ref{Fig.2 ELF}). As seen from the figures obtained using GGA, the region with the red color reveals higher electron localization while the blue color displays the regions with lower electron localization [see Fig \ref{Fig.2 ELF} (a-b) \& Fig \ref{Fig.2 ELF} (e-f)]. Whereas, the representation of colour coding is just the opposite for mGGA as compared to GGA [see [Fig. 2 (c-d) \& Fig \ref{Fig.2 ELF}(g-h)]]. The charge localizations are more in Cl- and Br-atoms due to their higher level of electronegativity than Li and Sn atoms. Thus, these findings support the results of the Mulliken population analysis. 
\par To comprehend the results of our obtained MPA and ELF, we have calculated the 2D and 3D-schematic plots of the Electron Difference Density (EDD)\cite{Steiner1982}\cite{Geleta2024} of the investigated compounds and are depicted in S1.

\subsection{Electronic Properties}
\begin{figure*} [hbt!]
	\centering
	\includegraphics[height=8cm, width=8cm]{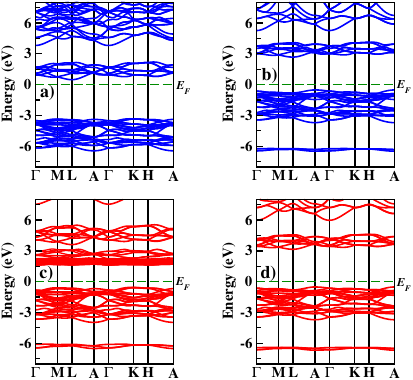}
	\includegraphics[height=8cm, width=8cm]{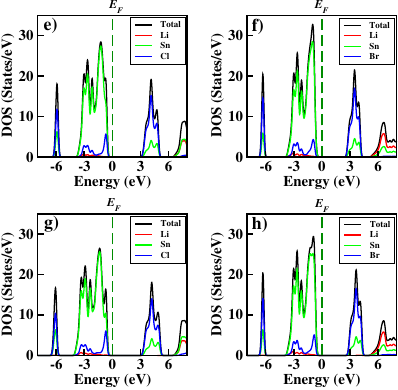}
	\caption{Calculated Band Structures and Density of States for LiSnX$_3$: (a,b and e,f ) using GGA, and (c,d and g,h) using mGGA}
	\label{Fig.3 Band-DOS}
\end{figure*}

While investigating atomic-level interactions in any material, one of the most significant aspects to take into account is its electrical properties.\cite{Chhana2022}\cite{Lalroliana2023b} In this study, we have employed GGA and mGGA functionals to analyze the electronic properties of LiSnCl$_3$ and LiSnBr$_3$. As presented in Table \ref{Table 2} and Fig \ref{Fig.3 Band-DOS}, the investigated materials exhibit indirect bandgaps with wide bandgaps semiconducting in nature. For both the employed functionals, the compounds containing Cl-atoms have wider \textit{E$_g$} than the Br-atoms containing compounds. This is due to the higher electronegativity level of the Cl-atoms.\cite{Das2022c}\cite{Wiktor2017a} For all the band structures, we considered the energy band range from -8 to +8 eV with the zero-point as the Fermi energy level (\textit{E$_F$}). Since all the under studied compounds are showing indirect bandgaps, we further observed that the maxima of valence bands lie on the high symmertry \textit{A}-points and the minima of conduction bands lie on the high symmetry \textit{L}-points. Here, all the studied compounds exhibit p-type semiconducting nature except for X=Cl (GGA) that has the nature of n-type semiconductor, which has higher conductivity and free electrons. Basically, it is important to note that the bandgap value drops gradually as halogen atoms are substituted by Cl$\rightarrow$ Br.\cite{Miah2024} The obtained \textit{E$_g$} are then compared with the existing results of LiSnCl$_3$ and, therefore, can be considered our values are in good agreement with the data obtained by others.\cite{Park2019,Lamichhane}

\par In Fig\ref{Fig.3 Band-DOS}, we have considered the energy range from -7 eV to 7 eV. The amplitude of the density of the states (DOS) is found to be ~30 eV. The DOS depicts the contribution level of the atomic orbitals, here, their valence orbitals. \cite{Lalhumhima2024}For A-site, the s-orbital of Li-atom is taken and for B-site, the p-orbital of Sn-atom has been taken. Furthermore, the p-orbital of experiment Cl/Br-atoms for X-sites have been taken into consideration. The contributions of the halogen atoms are higher in the valence bands and the same goes for Sn-atom in the conduction bands. From the obtained data and figures, we can say that wide bandgap semiconductors are good to be used in various applications such as power battery capacitors, flexible energy harvester, sensitive mechanical sensing, optogenetics and so on due to its characteristics in transfer of photons, from valence to conduction bands, are not in a quick succession.\cite{Roccaforte2018} One advantage of WBG nature is that it slows down the transfer of charge which avoid excessive movement of charge that may lead to explosion or can affect the material's durability.\cite{Castellazzi2019}

\subsection{Optical Properties}
An interaction of a material with electromagnetic (EM) or light waves is determined by its optical characteristics.\cite{Renthlei2023d} Understanding a material's optical capabilities is essential to comprehend its electrical configuration because they offers vital information about the most fundamental characteristics governing the material's suitability for optoelectronic and photovoltaic applications.\cite{MoufidaKrimi2024} In order to understand how the systems react to solar and high-energy wave radiations, we have examined the precise optical characteristics of LiSnX$_3$, such as the dielectric functions, absorption coefficients, refractive indices, and optical conductivities up to 8 eV photon energies. The corresponding optical responses along the xx- and yy-axes are the same and overlap each other because of the isotropy of planar symmetry. But the response along the zz-axis were different owing to anisotropic symmetry. The measured optical parameters have been plotted along the xx-, and zz-axes.

\par Using the real ($\epsilon_{1}$) and imaginary ($\epsilon_{2}$) parts of the dielectric functions, we can obtain the various optical output such as - absorption coefficient ($\alpha$), refractive indices ($\eta$), optical conductivities and so on.

\par The expression for the complex dielectric functions is given as\cite{Rahaman2018b}
\begin{equation}
\epsilon = \epsilon_{1} + \epsilon_{2}
\end{equation} 
\par The real part ($\epsilon_{1}$) of the dielectric constant can be obtained from the imaginary part ($\epsilon_{2}$) of the dielectric constants utilizing the Kramer-Kronig's transformation given as\cite{Liu2024}
\begin{equation}
\epsilon_{1}(\omega) = 1 + \dfrac{2}{\pi} \int_{0}^{\infty} \dfrac{\epsilon_{2}(\omega^{'})\omega^{'}d\omega^{'}}{\omega^{'2} - \omega^{2}}
\end{equation}
\par The imaginary part  ($\epsilon_{2}$) of the dielectric constant which provides the relation between unoccupied-occupied eigenstates of the electronic band structures' wave functions is given as
\begin{equation}
\begin{split}
\epsilon_2(\omega)= \frac{\hbar^2 e^2}{\pi m^2 \omega^2}\sum_{nn'}\int_{k}d^3k\big|\big<\vec{k}n|\vec{p}|\vec{k}n'\big>\big|\\
\times\big[1-f(\vec{k}n)\big]\delta(E_{\vec{k}n}-E_{\vec{k}n'}-\hbar\omega) 
\end{split}
\label{Eq 4}
\end{equation}
\par where $|\vec{k}_n$\big> represents the eigen function value E$_{\vec{k}n}$ , $\vec{p}$ is the operator of the momentum,  and \textit{f}{($\vec{k}$n)} being the Fermi distributional function.
\par From the above two equations, we can obtain other optical properties such as
\par The absorption coefficient ($\alpha$) of the materials can be obtain using 
\begin{equation}
\alpha(\omega)= \frac{2\omega k(\omega)}{c}
\label{Eq 5}
\end{equation}
\par  ${k(\omega)}$ being the extinction coefficient
\begin{equation}
k (\omega) =\sqrt{\frac{(\epsilon_1^2 + \epsilon_2^2)^\frac{1}{2} -\epsilon_1}{2}}
\label{Eq 6}
\end{equation}
\par The refractive index ($\eta$) determines the speed of light in a given substance and can be written as
\begin{equation}
\eta (\omega) = \sqrt{\dfrac{(\epsilon_1^2 + \epsilon_2^2)^\frac{1}{2}+\epsilon_1}{2}}
\label{Eq 7}
\end{equation}
\par The complex optical conductivity ($\sigma$) which connects to the complex dielectric constant ($\epsilon$) via the following equation.
\begin{equation}
\sigma_{1} = \omega\epsilon_{2}\epsilon_{0} \ , \  	\sigma_{2} = \omega\epsilon_{1}\epsilon_{0}
\end{equation}
\par where $\omega$ represents the angular frequency and $\epsilon_{0}$ is the free space dielectric constant.

\par Using GGA and mGGA, we have calculated the optical properties for LiSnCl$_3$ and LiSnBr$_3$. All the calculated properties have been measured upto +8 eV photon energy. The effectiveness of optoelectronic devices can be measured by the dielectric constants, which measure how efficiently a material responds to light. Thus, higher dielectric values result in greater performance of the optoelectronic devices.\cite{Zosiamliana2022i}
\par
\begin{figure}[hbt!]
	\centering
	\includegraphics[height=5cm]{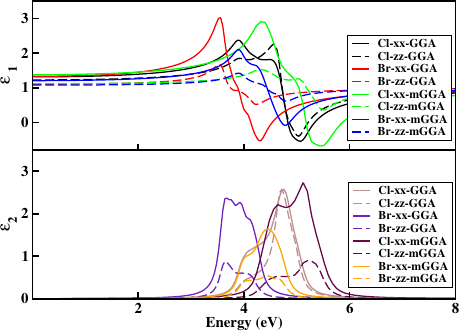}
	\caption{Dielectric Constants- Real \& Imaginary parts of LiSnX$_3$(X=Cl and Br) }
	\label{Fig.4 Optical-DC}
\end{figure}

 Since all the investigated materials are anisotropic in nature, the response of the optical properties at xx- and yy- directions overlap each other. Due to this, we have taken the directions of xx- and zz- in the following study. The real and imaginary parts of dielectric constants are depicted in Fig.\ref{Fig.4 Optical-DC}. We observed that the real parts ($\epsilon_{1}$) of LiSnCl$_3$ using GGA along the xx-axis exhibit its maximum value at E = 3.12 a.u. at 3.23 eV and attain its minimum value at E = -0.86 a.u. at 5.25 eV. Along its zz-direction, the highest value can be observed at E = 2.14 a.u with 4.84 eV and the minimum value reaches down to E = -0.65 a.u. at 5.21 eV. Using mGGA for X = Cl along the xx- and zz- direction, we obtained the value of E = 2.84 a.u. and 1.23 a.u. at 4.24 eV and 4.32 eV, respectively, and their minimum values drops at E = -0.91 a.u and 0.36 a.u at 5.32 eV and 5.31 eV. For the real parts ($\epsilon_{1}$) of LiSnBr$_3$ with employed GGA, the directions along xx- and zz- attain their highest values at E = 3.15 a.u and 1.27 a.u at 3.23 eV and 3.32 eV, respectively. Their minimum values are obtained at E = -0.75 a.u and 0.38 a.u at 4.22 eV and 4.12 eV, respectively. Using mGGA for X = Br, the maximum peaks along xx- and zz- directions are at E = 1.86 a.u and 1.22 a.u with the energy level of 3.85 eV and 3.87 eV, respectively. While their minimum value drops at E = -0.28 a.u and 0.28 a.u with energy level of 4.78 eV and 4.88 eV, respectively. The values of the real parts which drops below the zero-level are due to the phenomenon known as plasmonic vibration. Table.\ref{Table 4} shows the computed statics for real $\epsilon$($\omega$) and refractive $\eta$($\omega$) along the xx- and zz-axes.

\begin{figure}[hbt!]
	\centering
	\includegraphics[height=5cm, width=7cm]{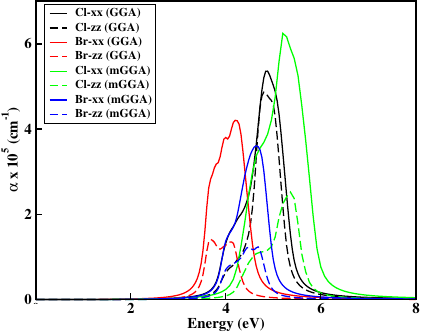}
	\caption{Absorption coefficients of LiSnX$_3$(X=Cl and Br) }
	\label{Fig.5 Optical-Absorp}
\end{figure}

\begin{figure}[hbt!]
	\centering
	\includegraphics[height=5cm]{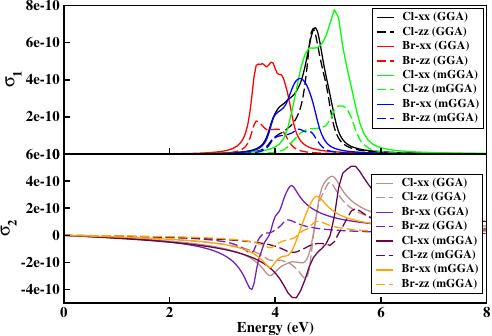}
	\caption{Optical conductivities of LiSnX$_3$(X=Cl and Br) }
	\label{Fig.6 Optical-Conductivity}
\end{figure}

\par The imaginary responses for X = Cl and Br are also obtained using GGA and mGGA. Here, the xx- direction response for X = Cl at their maximum peaks using GGA and mGGA levels is E = 2.75 a.u and 2.84 a.u at 4.75 eV and 5.16 eV, respectively. While their directions at zz- axis are E = 2.74 a.u and 0.54 a.u at 4.76 eV and 5.34 eV, respectively. Similarly, we have also calculated for X = Br where the xx- direction responses using both functionals are E = 2.33 a.u and 1.67 a.u at 3.51 eV and 4.56 eV, using GGA and mGGA. Likewise, for X = Br at zz-directions, the responses are at E = 0.85 a.u and 0.84 a.u at 3.64 eV and 5.23 eV, respectively for the employed GGA and mGGA. Herein, the imaginary parts of the dielectric constants, all the responses do not falls under the zero-level.

\begin{table}[hbt!]
	\small
	\caption{\ Calculated static real  $\epsilon_1$$\big(0\big)$ part of the dielectric constant and static refractive indices $\eta$$\big(0\big)$ along the xx-, zz-axes of LiSnX$_3$ (X = Cl and Br)}
	\label{Table 4}
	\begin{tabular*}{0.48\textwidth}{@{\extracolsep{\fill}}l|l|llll}
		\hline
		\textbf{X}& Functionals &$\epsilon_1$$^{xx}$$\big(0\big)$ & $\epsilon_1$$^{zz}$$\big(0\big)$  & $\eta$$^{xx}$$\big(0\big)$ & $\eta$$^{zz}$$\big(0\big)$\\
		\hline
		\textbf{Cl}&GGA & 1.26  & 1.25 &  1.22 & 1.13\\
		
		\textbf{Br}&GGA &1.28 & 1.28 & 1.82 &1.07\\
		
		\textbf{Cl}&mGGA  & 1.31 &1.31 &  1.24 &1.08\\
		
		\textbf{Br}&mGGA & 1.12 &1.12& 1.19 & 1.10\\
		\hline
	\end{tabular*}
\end{table}

\begin{table*}[hbt!]
	\small
	\caption{\ Calculated elastic constants (C$_{ij}$ in GPa), Cauchy pressure (in GPa), Kleimann coefficient ($\zeta$), eigenvalues ($\lambda_i$ in GPa), Density ($\rho$ in g/m$^3$), Vicker's hardness ($H_v$), Anisotropy ($A_{an}$),  and machinability index ($\mu_m$). }
	\label{Table 5}
	\begin{tabular*}{\textwidth}{@{\extracolsep{\fill}}llllllllllll}
		\hline
		\textbf{X}& Functionals & C$_{11}$ & C$_{12}$ & C$_{13}$ & C$_{22}$ & C$_{33}$ & C$_{44}$ & C$_{55}$ & C$_{66}$ & (C$_{12}$ - C$_{44}$) & $\zeta$
		\\\hline
		\textbf{Cl}& GGA & 27.76 & 10.60 & 2.49 & 27.65 & 22.55 & 2.11 &2.26 &8.58 & \ \ \ 8.49 & 0.51\\
		\textbf{Br}&GGA & 22.32 & 8.67 & 1.90 &  22.04 & 17.74 & 1.20 & 1.31 & 6.77 & \ \ \ 7.47 & 0.52\\
		\textbf{Cl}& mGGA & 28.77 & 14.18 & 3.55 & 32.58 & 25.73 & 6.02 & 3.01 & 10.56 & \ \ \ 8.16 & 0.61\\
		\textbf{Br}& mGGA & 13.31 & 8.82 & -1.28 & 19.03 & 16.81 & 2.04 & 1.05 & 5.59 &\ \ \ 6.78 & 0.76\\
		\hline
		&& $\lambda_1$ & $\lambda_2$ & $\lambda_3$ & $\lambda_4$ & $\lambda_5$ & $\lambda_6$ & $\rho$ & \textit{H$_v$} & \textit{A}$_{an}$ & \textit{$\mu$}$_m$
		\\\hline
		\textbf{Cl}& GGA & 1.14&1.28  &9.59 & 18.03 & 21.86& 38.99 & 2.96 & 0.52 & 2.47 & 5.60 \\
		\textbf{Br}&GGA & 0.51 & 0.68 & 7.44 & 14.15 & 17.27 &31.33 & 6.04 &0.35 & 2.46 & 7.80 \\
		\textbf{Cl}& mGGA &1.65 & 4.99 & 11.32 & 17.33 & 25.27 & 46.12 & 3.85 & 0.91 &2.43 &2.27\\
		\textbf{Br}& mGGA & -3.64 & 2.18 & 5.18 & 7.49 & 17.39 & 27.11 & 6.05 & 0.89 &1.63  & 4.30 \\
		\hline
		
	\end{tabular*}
\end{table*}

\par Our acquired absorption outputs in Fig.5 showed that the studied materials all falls within the range of the high-end of vis-UV region. As can be seen from the figure, among all the investigated compounds, X = Br (GGA) has the lowest peak within the vis-UV region. While X = Cl (mGGA) has the highest peak along the UV-region. Thus, these results refer to the materials are potential candidate for not only optoelectronics but also in the field of industrial processes and medical practices 

\par When referring to conducting medium in particular, optical conductivity amplifies electrical transport to high frequencies. There are two parts of optical conductivity- Real and imaginary parts. In Fig.\ref{Fig.6 Optical-Conductivity}, the real part ($\sigma_{1}$) corresponds with the $\epsilon_{2}$ and explains about the absorption with X = Cl (mGGA) at energy level 5.25 eV and X = Br (GGA) with the nearest range of vis-UV region. The imaginary parts ($\sigma_{2}$) aligns with the $\epsilon_{1}$ and deals with the level of the material's polarization.
\par The findings of the refractive index for these novel compounds are illustrate and reported in S2.

\subsection{Elastic constants and Mechanical Properties}
Mechanical stability is necessary for a material to be understood in its actual form. Using the finite-strain theory\cite{Weaver1976}, we may determine the compound's mechanical stability by computing the elastic constants (C$_{iv}$). For a material's hexagonal symmetry to be mechanically stable, it requires to fulfill the following criteria\cite{Yu2018b}
\begin{equation}
C_{44} > 0, C{_{11}^{2}} > C{_{12}^{2}}, (C_{11} + 2C_{12})C_{33} > 2C{_{12}^{2}}
\label{Eq 9}
\end{equation}

\par  All the calculated elastic constants of the under investigated halide perovskites presented in Table \ref{Table 5} fulfills the Born stability criterion\cite{Born1940b} of hexagonal symmetry as given in the above Equation \ref{Eq 9}. Herein, we observed that from Table \ref{Table 5}, the elastic constants of C$_{11}$, C$_{22}$, and C$_{33}$ are significantly greater than C$_{44}$, C$_{55}$, and C$_{66}$ for all the studied compounds under the employed functionals- GGA and mGGA. This suggests the materials under study have higher resistance to axial compression than shear deformation which signifies the bulk modulus (B) is greater than that of the shear modulus (G) as can be seen in Table \ref{Table 6}. Despite being fulfilling the Born stability criterion using the elastic constants, the studied compounds' eigen-state values are all positive except for LiSnBr$_3$ (mGGA), which means that it cannot achieve a pure form of crystalline stability. The Kleinman coefficient ($\zeta$) whose value ranges between 0$\le$ $\zeta$ $\le$ 1, depicts a material's resistance to bending and stretching have been utilized to study the internal contraction stability using the equation\cite{Kleinman1962a}

\begin{equation} 
\zeta = \dfrac{C_{11} + 8C_{12}}{7C_{11} + 2C_{12}}
\label{Eq 10}
\end{equation}
\par The $\zeta$-values closer to one ($\zeta$ $\rightarrow$ 1) indicates a higher contribution towards bond bending and the value tends to zero ($\zeta$ $\rightarrow$ 0) implies higher contribution towards bond stretching for each compound. From our obtained results displayed in Table \ref{Table 5}, it is possible to state that the mechanical toughness of the examined compounds increases likely towards bond bending. From our observations, the compounds under GGA are more likely to have bond stretching nature than the compounds under employed mGGA functional which shows more of bond-bending nature. We have also observed that X = Cl and Br under GGA can act as an intermediate contributors for both bending and stretching.

\par The calculated elastic moduli obtained in terms of Voigt, Reuss, and Hill's assumptions are shown in Table \ref{Table 6}. The elastic moduli derived from elastic constants follow the Born stability requirements and are mechanically stable for all examined materials. The Bulk (B), Young's (E), Shear (G) moduli all are important elastic constants which can derive the mechanical properties of the compounds. As a result of their low bulk (B) and shear (G) moduli, these compounds can be considered to be flexible in nature and could be comply for produce in thin films which makes them appropriate for optoelectronics applications.\cite{Shi2017a}

\begin{table*}[hbt!]
	\small
	\caption{\ Calculated values of elastic moduli - Bulk modulus(\textit{B}), Young's modulus(\textit{E}), Shear modulus(\textit{G}) all in GPa units, Pugh's ratio(\textit{k}) and Poisson's ratio(\textit{v}) (unitless). Here, the subscripts \textit{V}, \textit{R} and \textit{H} represent Voigt, Reuss and Hill assumptions, respectively}
	\label{Table 6}
	\begin{tabular*}{\textwidth}{@{\extracolsep{\fill}}lllllllllllllllll}
		\hline
		\textbf{X} &Functionals& \textit{B}$_V$ & \textit{B}$_R$ & \textit{B}$_H$ & \textit{E}$_V$ & \textit{E}$_R$ & \textit{E}$_H$ & \textit{G}$_V$ & \textit{G}$_R$ & \textit{G}$_H$ & \textit{k}$_V$ & \textit{k}$_R$ & \textit{k}$_H$ & \textit{v}$_V$ & \textit{v}$_R$ & \textit{v}$_H$\\
		\hline
		\textbf{Cl} &GGA& 12.08 & 11.53 & 11.80 & 17.10 & 7.00 & 12.29 & 6.76 & 2.51 & 4.64 & 1.79 & 4.61 & 2.55& 0.26 & 0.39 & 0.33\\
		\textbf{Br} &GGA& 9.62 & 9.11 & 9.37 & 13.17 & 3.67 & 8.69 & 5.18 & 1.29 & 3.23 & 1.86 & 7.12 & 2.90& 0.27 & 0.43 & 0.35\\
		\textbf{Cl} &mGGA&14.26 & 13.05 & 13.65 & 20.96 & 12.15 & 16.68 & 8.35 & 4.52 & 6.43 & 1.71 & 2.89 & 2.12 & 0.26 & 0.34 & 0.29\\
		\textbf{Br} &mGGA& 6.63 & 10.94 & 8.78 & 10.49 & 16.23 & 13.36 & 4.24 & 6.48 & 5.36 & 1.56 & 1.69 & 1.64 & 0.24 & 0.25 & 0.25\\
		\hline
	\end{tabular*}
\end{table*}

\par Using the elastic moduli, we can derive the material's brittleness and ductility. The Poisson's ratio (\textit{v}) which uses strain dependence calculation have been implemented to define the material's failure mode. The \textit{v} values below(above) the critical point (0.26) are considered to be brittle(ductile).  Also, the Pugh's ratio \textit{(B/G)} which differentiate the nature of a material where its value is below(above) 1.75 are subjected to be brittle(ductile). Furthermore, using elastic constants of (C$_{12}$ - C$_{44}$) also known as the Cauchy pressure define the material's failure mode where the positive value shows the material's ductility and negative value indicates the material to be brittle.
\par Thus, in this particular instance of hexagonal symmetry framework. We report that all the examined lead-free halide perovskites are mechanically stable, and the compounds LiSnCl$_3$ for both GGA and mGGA along with LiSnBr$_3$ for GGA  demonstrated ductile in nature, but LiSnBr$_3$ for mGGA can be classified as brittle material since its values of both Pugh's and Poisson's ratios are below the critical points.
\begin{table*}[hbt!]
	\small
	\caption{\ Calculated values of longitudinal wave (V$_l$), transverse wave (V$_t$) and mean sound velocities (V$_m$). All waves and sound velocities are in m/s. Melting temperature (\textit{T$_m$}, $\pm$ 300) (in K), machinability index ($\mu$), Debye's frequency(\textit{V$_D$})($\times$ 10$^{12})$ in \textit{Hz} and Debye's Temperature($\theta$$_D$) in \textit{K}}
	\label{Table 7}
	\begin{tabular*}{\textwidth}{@{\extracolsep{\fill}}llllllll}
		\hline
		\textbf{X}&Functional & V$_l$ & V$_t$ & V$_m$ & \textit{T$_m$} &\textit{V$_D$}& $\theta$$_D$\\
		\hline
		\textbf{Cl}&GGA&2960&1250&1410&717.08&3.4&163.1\\
		
		\textbf{Br}&GGA&1500&1030&1120&684.93&2.8&134.3\\
		
		\textbf{Cl}&mGGA&2400&1290&1450&723.06&3.5&167.9\\
		
		\textbf{Br}&mGGA&1620&\ 940&1040&631.67&2.5&119.9\\
		\hline
	\end{tabular*}
\end{table*}

\subsection{Anisotropic Insight}
The Zener factor, also known as the elastic anisotropic factor, \textit{A.} This describes the bulk features of a crystal structure that depend on its crystalline orientation. Analysing a material's isotropy/anisotropy factor is crucial for understanding its atomic arrangement, direction, and morphology. A material is said to be isotropy when it has the value of unity (\textit{A} = 1), otherwise if the value is \textit{A$_{an}$}$>$ \textit{A} $<$\textit{A$_{an}$} then the material is likely to be anisotropic in nature. Using the below equation we can obtain the values for \textit{A$_{an}$}\cite{Zhao2020}\cite{Shah2023}

\begin{equation}
A_{an} =\dfrac{4C_{11}}{C_{11} + C_{33} - 2C_{13}}
\label{Eq 11}
\end{equation}

\par The computed \textit{A$_{an}$} values are shown in Table \ref{Table 5}. The study considers all lead-free halide perovskites, revealing their highly anisotropy. Various indices can be used to analyze the level of isotropy and anisotropy.

\subsection{Vicker's Hardness}
One of the important property to acquire a material's hardness, especially in the thin and macroscopic sections, is to obtain the Vicker's hardness (\textit{V$_H$}). The values of examined compounds' \textit{V$_H$} can be obtain from\cite{Dovale-Farelo2022}

\begin{equation}
H_v = \dfrac{(1-2\textit{v})E}{6(1+\textit{v})}
\label{Eq 12}
\end{equation}

\par The calculated values can be seen in Table \ref{Table 5}. From the values, among the studied compounds, LiSnCl$_3$ (mGGA) has the highest \textit{V$_H$} value with 0.91. This occurs because the atom's size increases and the nucleus's attraction to the outermost electron lessens as we move down the group.

\subsection{Machinability Index}
In order to provide information about a material to be utilized in industries and commercialization, an index of machinability \textit{($\mu$$_m$)} can be calculated. The machine tool durability, hardness, capacity, cutting form, and other parameters affect this parameter. This factor can be described by the ratio of bulk modulus (B) to shear resistance (C$_{44}$).\cite{Ahmed2023a}

\begin{equation}
\mu_m = \dfrac{\textit{B}}{C_{44}}
\label{Eq 13}
\end{equation}

\par An index $\mu$$_m$ of machinability greater than 1.45 is considered to be good in manufacturing. The obtained values presented in Table \ref{Table 5} shows that all the lead-free perovskites under considerations are all applicable for device fabrication.

\subsection{Melting Temperature}
Understanding a material's melting point is essential for real-world applications. In addition to simulating atomic and ionic radii, temperature-parameter is crucial for structural stability. Using the following equation, we have obtained the melting temperatures (\textit{T$_m$}) of the studied compounds\cite{Celestine2024b}\cite{Jung2023}

\begin{equation}
T_m = 553 + 5.911C{_{11}} \pm 300
\label{Eq 14}
\end{equation}
\par As shown in Table \ref{Table 7}, the calculated \textit{T$_m$} values using GGA and mGGA are aligned, where LiSnCl$_3$ in both functionals have higher melting temperatures than LiSnBr$_3$. Among them X = Cl (mGGA) has the highest value up to 723.06 $\pm$ 300 K which simply means that it can out-stand its structural integrity even at high temperatures comparing with the others' values.

\subsection{Sound Velocity}
Determining sound velocities is one of the important parameters when estimating mechanical properties. They can be obtain by using the symmetry of crystal and the direction of propagation. The longitudinal velocities,  \textit{v}$_\textit{l}$ =  $[(B + 4G/3)/\rho]^{1/2}$ and transverse velocities,  \textit{v}$_\textit{t}$ = (G/$\rho$)$^{1/2}$, also know as the seismic-compression and shear waves are used to obtain the mean sound velocities \textit{v}$_\textit{m}$. \cite{Rahman2024}

\begin{equation}
\textit{v}_{m} = [(1/\textit{v}^3_p + 2/\textit{v}^3_s )/3]^{-1/3}
\label{Eq 15}
\end{equation}

\par Our calculated results for \textit{v}$_\textit{l}$, \textit{v}$_\textit{t}$ and \textit{v}$_\textit{m}$ are presented in Table \ref{Table 7}. We observed that among the studied compounds, X = Cl (GGA) has the highest \textit{v}$_\textit{l}$ value while X = Cl (mGGA) has the highest \textit{v}$_\textit{t}$ value. The highest mean sound velocities is obtained by X = Cl(mGGA) with the value of 1450 m/s.
\par The true forms of hexagonal crystals can only be obtained in [100] and [001] directions. These can be derive using the following equations\cite{Celestine2024c}

\begin{equation}
\centering
\begin{split}
[100]:[100]\textit{v}_l= \sqrt{(C_{11} - C_{12})/2\rho}\\ 
[010]\textit{v}_{t1} = \sqrt{C_{11}/\rho}\\ 
[001]\textit{v}_{t2}= \sqrt{C_{44}/\rho}
\end{split}
\centering
\label{Eq 16}
\end{equation}

\begin{equation}
\centering
\begin{split}
[001]:[001]v_l= \sqrt{(C_{33}/\rho}\\ 
[100]v_{t1} = \sqrt{C_{44}/\rho}\\ 
[010]v_{t2}= \sqrt{C_{44}/\rho}
\end{split}
\centering
\label{Eq 17}
\end{equation}   

\begin{table*}
	\small
	\caption{\ Calculated values of sound velocities for LiSnX$_3$ (X = Cl, and Br) along different directions [100] and [001].}
	\label{Table 8}
	\begin{tabular*}{\textwidth}{@{\extracolsep{\fill}}llllllll}
		\hline
		\textbf{X} & Functionals&\textbf{[100]} &  &  & \textbf{[001]} & & \\
		\hline
		&& [100]$v_l$ & [010]$v_{t1}$ & [001]$v_{t2}$ & [001]$v_l$ & [100]$v_{t1}$ & [010]$v_{t2}$ \\
		\hline
		
		\textbf{Cl} &GGA& 1.73 & 3.06 & 0.84 & 2.76 & 0.84 & 0.84 \\
		\textbf{Br} &GGA& 1.12 & 1.92 & 0.45 & 1.71 & 0.45 & 0.45\\
		\textbf{Cl} &mGGA& 1.37 & 2.73 & 1.25 & 2.59 & 1.25 & 1.25 \\
		\textbf{Br} &mGGA& 0.61 & 1.48 & 0.58 & 1.67 & 0.58 & 0.58\\
		\hline
	\end{tabular*}
\end{table*}

\par Utilizing the above two equations \ref{Eq 16} and \ref{Eq 17}, we obtained the values of sound velocities for these compounds in the directions of x[100] and z[001], respectively. In the equations, the first and second transverse modes are represented by $v_{t1}$ and $v_{t2}$, respectively. As listed in Table \ref{Table 8}, along [100] propagation direction, X= Cl (GGA) has the highest sound velocity at [010]$v_{t1}$ polarization direction while the smallest value results along [001]$v_{t2}$ polarization direction of X = Br (GGA). Along [001] propagation direction, again X = Cl (GGA) tops the sound velocity and the polarization directions along [100]$v_{t1}$ and [010]$v_{t2}$ possess identical sound velocities. Thus, in the study of sound velocities, GGA has more impact on these studied compounds.
\subsection{Debye's Temperature and Frequency}
In material science, Debye temperature\textit{($\theta$$_{D}$)} is a crucial metric that aids in understanding how materials behave under severe heat and cold. It also sheds light on the mechanical and thermal characteristics of materials. The \textit{$\theta$$_{D}$} also indicates the toughness of a material and can even dictates the number of active phonon modes through its relationship with sound velocities, density and mass of the material. This phenomenon can be activated by the crystal's structure, chemical composition and other external factors like pressure and temperature. Using the following equation we can obtain the Debye's temperature of the materials\cite{Rahman2024}

\begin{equation}
\theta_{D} = \dfrac{\textit{hv$_{D}$}}{k}
\label{Eq 18}
\end{equation}

\par where \textit{h} is the Planck's constant, {\textit{v$_{D}$} is the Debye's frequency and \textit{k} denotes the Boltzmann constant.
	\par In order to acquire to the value of \textit{$\theta$$_{D}$}, first we need to have the debye's frequency {\textit{v$_{D}$}, which solely depends on the number of atoms per unit volume and the sound velocities. The equation given below can be utilized to obtain the debye's frequency\cite{Chen2001}\cite{Gan2018}
		
		\begin{equation}
		\textit{v$^{3}$$_{D}$} = 2\pi. 9(N/V)  \Bigg[\dfrac{2}{v^{3}_{t}} + \dfrac{1}{v^{3}_{l}}\Bigg]^{-1}
		\label{Eq 19}
		\end{equation}
		
		\par Using Equations \ref{Eq 18} and \ref{Eq 19}, we calculated the values of Debye's temperatures and frequencies of the lead-free halide compounds which are presented in Table \ref{Table 7}. We observed that  X = Cl (mGGA) has the highest values of 	\textit{v$_{D}$} and 
		$\theta_{D}$ which signifies that it has the best mechanical attributes and can out-stand the temperal conditions as compared to the others.
		\subsection{Piezoelectric Properties}
		Piezoelectric properties have sparked significant study attention due to their ability to convert green energy.\cite{Singh2023a}\cite{Li2022} Its effect comes into action when mechanical stress is applied to an asymmetric crystal, causing atomic polarisation of positive and negative ions from their starting positions.\cite{Li2022} Larger polarisation results in enhanced reaction to piezoelectric response.\cite{King-Smith1993a} A material's electrical and mechanical states can be coupled by piezoelectric, which depends on the lattice structure of either naturally occurring quartz or artificial materials that mimic quartz.\cite{Ippili2021} It has recently been reported that halide perovskites, because of their large band gaps and non-centrosymmetric structures, have shown a strong response to piezoelectricity.\cite{Young2016}\cite{Chakraborty2023} Furthermore, it is believed by researchers as well as companies to be a significant advancement that replacing the lead (Pb) content in the field of biomedical applications eliminates potential damage and lowers toxicity.\cite{Collin2022}\cite{Wu2022b}\cite{Mallick2021}\cite{Wu2022c}
		
		\par Determining the investigated halide perovskites' piezoelectric tensors is the primary goal of computing their piezoelectric properties. Piezoelectric materials show direct electric polarisation when an external macroscopic strain is applied, and this polarisation can become inverse by creating an external electric field. There are various ways where piezoelectric materials can be utilized as storage system for energy and in other applications- microelectromechanical systems (MEMS).\cite{Bassiri-Gharb2008}\cite{Ge2023} Due to the shortage of reliable details or conclusions regarding the researched halide perovskites' piezoelectric properties, we then calculated the piezo-tensors for these compunds. Along with these, using the calculated elastic constants, dielectric constants, permittivity of space and the piezo-tensors, we obtained the electromechanical coupling constants to verify the efficiency and electrical conversion rates of these lead-free halide perovskites.
		\par The total macroscopic polarisation \textit{(P)} of a bulk system, in the absence of external field interference, is the combination of the strain-dependent piezoelectric polarisation produced by strain \textit{P$_{p}$} and the spontaneous polarisation \textit{P$_{eq}$} of the equilibrium structure given as\cite{Bernardini1997a}
		\begin{equation}
		P = P_{p} + P_{eq}
		\label{Eq 20}
		\end{equation}
		\par The piezo-tensors can 
		be expressed in terms of
		\begin{equation}
		\gamma_{\delta\alpha} = \dfrac{\triangle P_\delta}{\triangle \in_\alpha}
		\label{Eq 21}
		\end{equation}
		
		\par where $\gamma_{\delta\alpha}$ can be obtained using the QuantumATK code which uses the finite- difference method and the polarisation \textit{P} can be obtained employing the Berry-phase approximation method.\cite{Rohrlich2009a}\cite{Bell1964a} A different approach to determining the piezoelectric tensors requires setting up two terms: 
		\\i) the clamped-ion term e$_{i,j}$, which represents the electronic response to strain; \\
		ii) a term that define how the internal strain affects the piezoelectric polarisation.
		\par Consequently, the full expression for e$_{i,j}$ is obtained as 
		
		\begin{equation}
		e_{ij} = e_{ij}(0) + \dfrac{4eZ^*}{\sqrt{3}a^2} \dfrac{du}{d{\in_\alpha}}
		\label{Eq 22}
		\end{equation}
		
		\par where \textit{i} denotes the direction along the current is applied and \textit{j} represents the strain directions. Z is the Born effective charge, which depends on the change in polarisation following the displacement of an ion, and \textit{e} is the electronic charge. $\in_\alpha$ is the macroscopic applied strain. 
		
		\begin{table}[h]
			\small
			\caption{\ Calculated piezoelectric tensors for the halide provskites LiSnX$_3$ (X=Cl, and Br) with different strains along x-, y- and z- directions under different functionals.}
			\label{Table 9}
			\begin{tabular*}{0.48\textwidth}{@{\extracolsep{\fill}}l|l|l|lll}
				\hline
				\textbf{X} &Functionals& Strains & \ \ \ \ x & \ \ \ \ y & \ \ \ \ z \\
				\hline
				\centering
				&& xx & \ 1.66e-01   & \ 8.79e-04 & -6.59e-02\\
				&& yy	& -1.66e-01 & \ 8.79e-04 & -6.59e-02\\
				&& zz & -3.14e-11 & -2.46e-03 & -2.59e-01\\
				\textbf{Cl}&GGA& yz	& -1.58e-01  & -1.24e-01& \ 9.43e-05\\
				&&	xz  & -1.58e-01    & \ 1.24e-01  & -9.43e-05\\
				&&	 xy & \ 7.37e-11 & -1.47e-01 & -2.22e-04\\
				
				\hline
				&& xx & -3.06e-02   & \ 2.19e-02 & \ 1.58e-01\\
				&& yy	& \ 3.07e-02 & \ 2.19e-02 & \ 1.58e-01\\
				&& zz & -6.36e-13 & -4.60e-04 & \ 2.81e-01\\
				\textbf{Br}	&GGA& yz	& \ 2.21e-01  & \ 1.80e-01  & \ 5.24e-02\\
				&&	xz  & \ 2.21e-01    & -1.80e-01  & -5.24e-02\\
				&&	 xy & -4.19e-13 & -6.40e-02 & \ 3.37e-02\\
				
				\hline
				&& xx & -1.66e-01    & \ 8.79e-04 & -6.59e-02\\
				&& yy	& -1.66e-01 & \ 8.79e-04 & -6.59e-02\\
				&& zz & -3.14e-11 & -2.46e-03 & -2.59e-01\\
				\textbf{Cl}&mGGA& yz	& -1.58e-01  & -1.23e-01& \ 9.43e-05\\
				&&	xz  & -1.58e-01  & \ 1.23e-01 & -9.43e-02\\
				&&	 xy & \ 7.37e-11 & -1.47e-01 & -2.22e-04\\
				\hline
				
				&& xx & \ 1.47e-01   & -4.03e-04 & -1.07e-01\\
				&& yy	& -1.47e-01 & -4.03e-04 & -1.07e-01\\
				&& zz & -1.69e-10 & \ 4.58e-04 & -1.59e-01\\
				\textbf{Br}&mGGA& yz&-1.56e-01  & -1.18e-01  & -1.42e-04\\
				&&	xz  & -1.56e-01    & \ 1.18e-01  & \ 1.42e-04\\
				&&	 xy & \ 1.10e-11 & -1.27e-01 & \ 3.76e-05\\
				\hline
			\end{tabular*}
		\end{table}
		
		\par The calculated piezoelectric tensors for the eco-friendly compounds are hereby listed in Table \ref{Table 9}. Here, we obtained 18 piezo-tensors by each individual lead-free halide systems which composed of 6 different strains namely: xx-, yy-, zz-, yz-, xz-, and xy- strains along x-, y-, and z- directions.  We can see from the table that there are six direct effects and twelve inverse effects for X= Cl (GGA) in various directions. There are further 11 direct and 7 inverse effects for X = Br (GGA). Moreover, 5 direct and 13 inverse effects were found for X = Cl (mGGA). Lastly, there are six direct and twelve inverse piezoelectric effects for X = Br (mGGA).
		
		\par We learnt that all the highest piezo-response for each system from the reported tensors, for X = Cl(GGA), Br(GGA), Cl(mGGA), and Br(mGGA) all originated from the corresponding z-directions along zz-strains with their response coefficients of -0.25 Cm$^{-2}$, 0.28 Cm$^{-2}$, -0.25 Cm$^{-2}$, and -0.15 Cm$^{-2}$, respectively. It is clear from the tensors report that X = Br(GGA) has the best piezoelectric tensors of all the compounds under study because it has more electrons in its atomic orbital, which increases its response when polarisation occurs as a result of an external force acting on it. Here, the direct and inverse effects indicate the strain directions in positive or negative directions. The obtained values of the investigated piezoelectric response are considerably higher than the average $\alpha$-quartz piezo-response coefficients (0.171 Cm$^{-2}$) reported by Bechmann et al.\cite{Bechmann1958b} and the temperature-dependent values of the same $\alpha$-quartz (0.07 Cm$^{-2}$) found by Tarumi et al.\cite{Tarumi2007a} Our studied halide perovskites exhibited a higher piezoelectric responsiveness (0.101 Cm$^{-2}$) than the mechanically superior glass-like Na$_2$SiO$_3$ materials reported by R. Zosiamliana et al.\cite{Zosiamliana2022l} As a result, the halide perovskites under study could be employed as piezoelectric or ferroelectric device materials.
		
		\subsection{Electromechanical Coupling Constant}
		The efficiency with which a piezoelectric material transforms electrical energy into mechanical energy or mechanical energy into electrical energy is measured by the electromechanical coupling factor ({k$_{ij}$}). Based on the supplier of piezoelectric standards, the coupling factors indicate the highest values that can be obtained in the theoretical approach. Employing the given equation, we can calculate the efficiency of the coupling factors\cite{Roy2012b}
		\begin{equation}
		\textit{k}_{ij} = \dfrac{|\textit{e}_{ij}|}{\sqrt{C_{ij}\epsilon_{fs}\epsilon_0}}
		\label{Eq 23}
		\end{equation}
		
		\par where k$_{ij}$ is the electromechanical coupling coefficient, $\textit{e}_{ij}$ being the calculated piezoelectric coefficient (Cm$^{-2}$), C$_{ij}$ indicates the elastic constant (10$^{11}$ Pa), $\epsilon_{fs}$ denotes the permittivity of space and $\epsilon_0$ being the static dielectric constant at zero-stress level. To carry out a linear-response calculation to obtain the electronic dielectric constants we use perturbed density functional theory (DFT) as implemented in QuantumATK.\cite{Smidstrup2019a}\cite{Otto1992a}	
		\begin{table}[hbt!]
			\small
			\caption{\ Calculated values of the electromechanical coupling constants for CsGeX$_3$ (X=Cl, Br and I) where e$_{ij}$ represents the piezoelectric coefficients and k$_{ij}$ being the electromechanical coupling coefficient.}
			\label{Table 10}
			\begin{tabular*}{0.48\textwidth}{@{\extracolsep{\fill}}l|l|ll}
				\hline
				\textbf{X} &Functionals& \ \textit{e$_{ij}$} & \textit{k$_{ij}$} \\
				\hline
				\textbf{Cl}     &GGA& -0.25   &0.061 \\
				\textbf{Br}     &GGA& \ 0.28   &0.063\\
				\textbf{Cl}     &mGGA&-0.25   & 0.046 \\
				\textbf{Br}     &mGGA& -0.15   & 0.037\\
				\hline
			\end{tabular*}
		\end{table}
		
		\par In Table \ref{Table 10} we provide the electromechanical coupling constants by taking the most significant piezoelectric constants from each of the lead-free halide perovskites that were examined. Among the studied compounds, X = Br(GGA) has the highest piezo-electromechanical response.
		
		\section{Environmentally Friendly Perovskites}
		The basic features of the investigated anisotropic structures of lead-free metal halide perovskites under the employed two functionals are shown in Table \ref{Table 11}. Herein, we observed that all the optical constants are within the high-energy (Vis-UV) regions. Within the low-energy region X = Br(GGA) has the highest optical constant while within the region of high-energy, X= Cl(mGGA) has the highest constant which also implies that the materials are applicable for optoelectronic devices. In contrast to those that contain lead (Pb) in their compounds, these ones have lower rigidity or bulkiness despite their stability criteria, but they may still be rendered and shaped into thin-film materials with less toxicity along with these because of their adequate electromechanical responses (piezo), which could also be used in piezoelectric devices.
		\begin{table}[hbt!]
			\small
			\caption{\ An outline of the main characteristics of the halide perovskites under investigation.}
			\label{Table 11}
			\begin{tabular*}{0.48\textwidth}{@{\extracolsep{\fill}}l|l|llll}
				\hline
				\textbf{X} &Functionals& Stability & Absorption &Fail mode& \\
				\hline
				\textbf{Cl}  &GGA& Stable* & High[UV-region] & Ductile&\\
				& & &   Low [Vis-region] &&\\
				\textbf{Br} &GGA&  Stable* & High[UV-region] & Ductile&\\
				& & & High-end[Vis-region]&&\\
				\textbf{Cl}  &mGGA& Stable* & High[UV-region] &Ductile&\\
				& & &   Low [Vis-region] &&\\
				\textbf{Br} &mGGA&  Stable* & High[UV-region] & Brittle&\\
				& & &  Low [Vis-region]&&\\
				\hline
			\end{tabular*}
			(Note = * indicates mechanical and octahedral)
		\end{table}

\section*{Conclusions}
	We comprehensively examined the characteristics of the novel halide-based perovskites LiSnX$_3$ (X= Cl and Br) using   GGA and mGGA functionals within the DFT formalism.  The ground state stabilities of the investigated compounds have been confirmed based on the elastic constants under born stability criteria and octahedral factors. We report a considerable agreement between the band gap calculated from mGGA and the previous existing theoretical data.  On analysis of the charge transfer, we have observed the accumulation of charge on the halogen (Cl/Br) atoms, whereas the charge depletion on the neighboring atoms. This mechanism led to the formation of ionic bonds. The halogen atoms contribute more in the valence band, while the B-site and Sn-atoms, contribute in the conduction bands. Along with this, we have also observed the n-type semiconducting behavior in X=Cl-based perovskite, promoting faster charge transfer.  The absorption peaks for both compounds falls within the  Vis-UV range, which indicates their optoelectronic functionality. The WBG semiconductors with high absorption coefficients within the UV-region are often used as a controller or modulator for UV-ray devices. An anisotropic dielectric polarization yields promising piezoelectric response in LiSnX$_3$. The electromechanical coupling constants are calculated by using the piezoelectric tensor constants. Thus, employing a novel approach and calculations, we acquired an insightful responses in optical and piezoelectric measures for these rarely studied compounds. Therefore, we hereby conclude that LiSnX$_3$ can be potential materials for technological applications.

\section*{Author contributions}
		\textbf{Celestine Lalengmawia:} Formal analysis, Visualization, Validation, Calculating results, Writing-original draft, writing-review \& editing.\\
		\textbf{Zosiamliana Renthlei:}Formal analysis, Visualization, Validation, writing-review \& editing. \\
		\textbf{Shivraj Gurung:} Formal analysis, Visualization, Validation, writing-review \& editing. \\ 
		\textbf{Lalhriat Zuala:} Formal analysis, Visualization, Validation, writing-review \& editing. \\ 
		\textbf{Lalrinthara Pachuau:} Supervision, Formal analysis, Visualization, Validation, writing-review \& editing. \\ 
		\textbf{Ningthoujam Surajkumar Singh:}Formal analysis, Visualization, Validation, writing-review \& editing. \\ 
		\textbf{Lalmuanpuia Vanchhawng}:Formal analysis, Visualization, Validation, writing-review \& editing. \\ 
		\textbf{Karthik Gopi}: Scripting, Formal analysis, Visualization, Validation, writing-review \& editing. \\
		\textbf{A. Yvaz}:Formal analysis, Visualization, Validation, writing-review \& editing. \\		\textbf{D.P. Rai:} Project management, Supervision, Resources, software, Formal analysis, Visualization, Validation, writing-review \& editing. \\ 

\section*{Conflicts of interest}
There are no conflicts to declare.

\section*{Data availability}

The data that support the findings of this study are available from the corresponding author, upon reasonable request.

\section*{Acknowledgements}
The research is partially funded by the Ministry of Science and Higher Education of the Russian Federation as part of World-class Research Center program: Advanced Digital Technologies (contract No. 075-15-2022-312 dated 20.04.2022)



\balance


\bibliography{rsc} 
\bibliographystyle{rsc} 

\end{document}